\documentstyle[prl,aps,multicol,epsf]{revtex}
\begin{document}
\draft
\title{Hall Coefficient of a Dilute 2D Electron System in Parallel
Magnetic Field.}
\author{S.~A.~Vitkalov, H. Zheng, K. M. Mertes and M.~P.~Sarachik}
\address{Physics Department, City College of the City
University of New York, New York, New York 10031}
\author{T.~M.~Klapwijk}
\address{Delft University of Technology, Department of Applied Physics,
2628 CJ Delft, The Netherlands}
\date{\today}
\maketitle \begin{abstract}
Measurements in magnetic fields applied at a small angle with respect to the 2D 
plane of the electrons of a low-density silicon MOSFET indicate that the Hall 
coefficient is independent of parallel field from $H=0$ to $H>H_{sat}$, the field 
above which the longitudinal resistance saturates and the electrons have reached 
full spin-polarization.  This implies that the mobilities of the spin-up and 
spin-down electrons remain comparable at all magnetic fields, and suggests there is 
strong mixing of spin-up and spin-down electron states.

\end{abstract}


\begin{multicols}{2}

Dilute, strongly interacting two-dimensional systems of electrons and holes have 
drawn intensive recent attention due to their anomalous behavior as a function of 
temperature and magnetic field\cite{rmp}: the resistance exhibits metallic 
temperature-dependence above a critical density, $n_c$, raising the possibility of 
an unexpected metallic phase in two dimensions \cite{krav}.  An additional 
intriguing characteristic of these systems is their enormous response to magnetic 
fields applied in the plane of the electrons \cite{simonian,pudalov} or holes 
\cite{cambridge,yoon}: the resistivity increases up to several orders of magnitude 
(depending on density, temperature, and the mobility of the sample), and saturates 
to a new value above a density-dependent characteristic magnetic field $H_{sat}$.  
Recent experiments \cite{okamoto,vitkalov} have shown that the field $H_{sat}$ 
corresponds to the onset of full spin polarization of the 2D electron system.  
With increasing parallel magnetic field the system thus evolves from zero net spin 
polarization, with equal numbers of spin-up and spin-down electrons, to a completely 
spin-polarized system.  Based on straightforward arguments, one expects different 
screening \cite{Dolgopol2} and different field-dependent mobilities for the spin-up 
and spin-down electrons.  The purpose of the Hall effect measurements reported in 
this paper was to find evidence for these two distinct sets of carriers.  
Our results indicate that the Hall coefficient does not, in fact, vary with a 
parallel magnetic field ranging from $0$ to well above $H_{sat}$.  Within a simple 
single-particle interpretation \cite{ziman} this indicates that, despite very 
different parameters such as Fermi velocities, the transport mobilities of the 
spin-up and spin-down electrons are comparable over the entire range of magnetic 
fields.  The fact that the Hall coefficient is independent of the degree of 
polarization of the electrons implies that frequent spin-flip scattering events 
cause substantial mixing of the spin-up and spin-down electron bands, highlighting 
the importance of electron-electron interactions in these dilute 2D electron systems.

\vbox{
\vspace{0.2in}
\hbox{
\hspace{-0.2in} 
\epsfxsize 3.3in \epsfbox{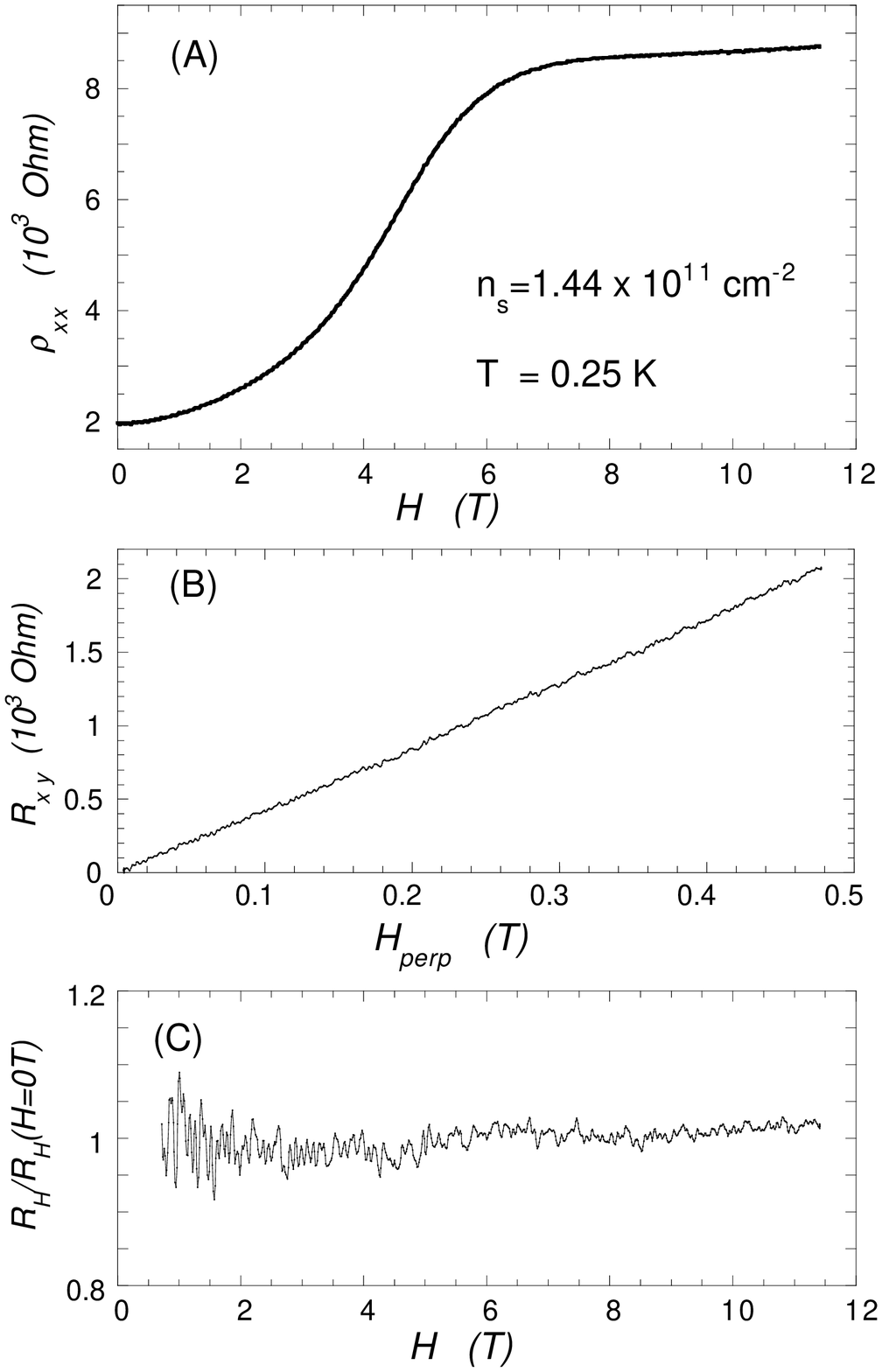} 
}
}
\refstepcounter{figure}
\parbox[b]{3.3in}{\baselineskip=12pt FIG.~\thefigure.
(A): Longitudinal resistivity versus in-plane magnetic field of a silicon 
MOSFET of electron density $n_s=1.44 \times 10^{11}$ cm$^{-2} \approx 1.7 n_c$, 
where $n_c$ 
is the critical density for the metal-insulator transition in zero field.  The 
temperature is $0.25$ K.  (B) The Hall resistance plotted as a function of the 
normal component of the magnetic field, $H_{\bot} = H$ sin $2.4^{\circ}$.  (C) The 
Hall coefficient, $R_H = R_{xy}/H_\bot$, as a function of parallel magnetic field.
\vspace{0.10in}
}
\label{1}

Measurements were performed on three silicon MOSFETs of comparable mobilities 
$\mu \approx 20,000\;$V/(cm$^2s)$ at $T=4.2$.  Contact resistances were minimized by 
using a split-gate geometry which allows a higher electron density in the vicinity 
of the contacts than in the 2D system under investigation.  Data were taken with 
standard four-probe techniques using $AC$ phase-sensitive detection to eliminate 
thermoelectric as well as other parasitic rectified signals.  Small unavoidable 
misalignment of the transverse Hall potential contacts introduces an admixture of 
the longitudinal resistance, which is known to be large and strongly field-dependent 
in these dilute 2D systems.  The longitudinal component was eliminated by averaging 
the voltages measured for two opposite directions of the magnetic field.  Data were 
taken in the linear regime with $AC$ currents typically below 5 nA, at 
frequency 3Hz.

The sample was mounted on a rotating platform at the end of a low temperature probe 
in a $^3$He Oxford Heliox system and measurements were obtained for temperatures 
between $0.235$~K and $1.66$ K in magnetic fields $H$ up to $12$ T.  The angle $\phi$ 
between the magnetic field $H$ and the 2D plane was determined by measuring the Hall 
resistance; $\phi$ was chosen to be very small in order to limit the size of 
the perpendicular field component to avoid quantum oscillations of the longitudinal 
resistance and Hall signal.

For a silicon MOSFET with density $n_s= 1.44 \times 10^{11}$ cm$^{-2}$, Fig. 1 (A) 
shows the longitudinal resistivity $\rho_{xx}$ at temperature $T=0.25$ K as a 
function of magnetic field applied at an angle $\phi = 2.4^\circ$; here 
$n_s \approx 1.7 n_c$, where $n_c$ is the critical density for the zero-field 
metal-insulator transition.  For an angle of $2.4^\circ$, the in-plane component 
of the magnetic field differs from the total field by only $0.1$\%, 
$H_{\|} = H$ cos $\phi = 0.999H$, while the perpendicular magnetic field remains 
small: $H_{\perp} = H$ sin $\phi = 0.0419 H$.  When the in-plane component has 
reached $12$~T the perpendicular component is only $0.5$ T; no Shubnikov-deHaas 
oscillations are apparent in either component of the resistivity, $\rho_{xx}$ or 
R$_{xy}$.  In agreement with earlier measurements for densities well above $n_c$, 
Fig. 1 (A) shows that the resistance increases by a factor of approximately four and 
saturates to a constant value above a magnetic field $H_{sat} \approx 6$ T.  Earlier 
measurements have shown that the resistance varies weakly with temperature for 
this electron density in a large magnetic field $H = 10.8$~T $>H_{sat}$, so that the 
sample has not entered the insulating phase in the highest field used in our 
experiments\cite{hairong}.  For the same sample under the same conditions, 
Fig. 1 (B) shows that the transverse Hall resistance is a clean linear function of 
perpendicular magnetic field up to $H_{\bot} = 0.5$ T, for in-plane magnetic 
field increasing to about $12$~T.  Fig. 1 (C) shows the Hall coefficient, 
$R_H = R_{xy}/H_{\bot}$, normalized to its value in zero field plotted as a function 
of in-plane magnetic field.  The Hall coefficient is constant to within $1-3$\%, the 
experimental error of our measurements, while the parallel component 
of the magnetic field changes from $0$ to well above $H_{sat}$.  The Hall 
coefficient is thus independent of the degree of polarization of the 2D electron 
system.

Temperature dependent screening of charged impurities 
\cite{Stern,Ando1,DasSarma2,Dolgop1} has been proposed as a possible explanation for 
the unusual metallic temperature dependence of the resistivity of dilute 2D 
systems.  Dolgopolov and Gold \cite{Dolgopol2} have recently argued that the 
screening properties of the electrons also depend strongly on in-plane magnetic 
field, causing the resistance to increase with field and to saturate when the 
electrons reach full spin polarization at $H>H_{sat}$, as observed experimentally 
\cite{okamoto,vitkalov}.  We note that the effect of screening on the behavior of 
the Hall coefficient has not been considered.  In what follows, we first discuss the 
dependence of the longitudinal resistance on magnetic field due to field-induced 
changes in screening and 
compare it with the measured magnetoresistance.  We then extend these concepts to 
consider the effect of field-dependent screening on the Hall coefficient.

In the Born approximation, the 
probability of electron scattering depends directly on the dielectric function  
$\epsilon$.  Due to the sharp edge of the electron distribution at $E=E_F$, the 
dielectric function $\epsilon (q)$ has a singularity at $q=2K_F$ in the limit $T=0$.  
Increasing the temperature causes smearing of the Fermi distribution, giving rise 
to a temperature-dependent resistance \cite{Stern}.  The effect of parallel field 
on electron screening was recently considered by 
Dolgopolov and Gold \cite{Dolgopol2}. A strong in-plane magnetic field decreases 
the energy of electrons with spins aligned along the magnetic field (spin-up 
electrons) and increases the energy of electrons with opposite spin (spin-down 
electrons).  The radius of the Fermi circle of spin-up (spin-down) electrons, 
$K_F^\uparrow$ ($K_F^\downarrow$), increases (decreases) with field.  The ability of 
the electron sea to screen the 
Fourier components of the external potential $V(q>2K_F^\downarrow)$ with wavevectors 
$q>2K_F^\downarrow$ is reduced because the spin-down particles with wavevector 
$K \leq K_F^\downarrow$ cannot interact with the potential $V(q)$ due to momentum 
conservation: $\vert \vec K^\downarrow_{fin}-\vec K^\downarrow_{in} \vert \leq 
2K_F^\downarrow < q$.  In other words, the short wavelength components of 
the external potential $V(q>2K_F^\downarrow)$ cannot be screened effectively by 
electrons with long de-Broglie wavelenghs.  The magnetic field thus reduces the 
screening of the short-wavelength components of the external potential by the 
spin-down electrons, causing an increase of scattering of the spin-up electrons.  
We thus expect the spin-up and spin-down electrons to have different mobilities 
in a magnetic field.

Results of calculations \cite {Dolgopol2} of the longitudinal resistivity as a 
function of in-plane magnetic field, performed in the clean limit 
($kT \gg\hbar/2\tau$), are compared with experiment in Fig. 2.  Although the shapes 
of the curves are different, due perhaps to approximations made in the calculations, 
there is reasonable agreement between theory and experiment in this range of 
electron densities.  We point out, however, that agreement between theory and 
experiment breaks down at lower densities near the metal-insulator transition, where 
the magnetoresistance is substantially 
larger (up to several orders of magnitude instead of a factor 4).

We now extend these concepts to consider the effect of screening on the Hall 
resistance.  As discussed above, the spin-up and spin-down electrons are expected to 
have different field-dependent mobilities in the presence of a parallel magnetic 
field.  If two 
distinct types of carriers with different, field-dependent mobilities contribute 
to the conductivity, the Hall coefficient $R_H$ is no longer expected to be constant 
as a function of magnetic field $H$.  For a two component ($\uparrow, \downarrow$) 
Fermi system the Hall coefficient is given by \cite{ziman}
$$
R_H=(\sigma_\uparrow \mu_\uparrow+\sigma_\downarrow \mu_\downarrow)/(\sigma_\uparrow  
+\sigma_\downarrow)^2,
$$   
where $\sigma_{\uparrow \downarrow}$ and $\mu_{\uparrow \downarrow}$ are 
the conductivities and mobilities, respectively, of the spin-up and spin-down 
electrons. 

Dolgopolov and Gold \cite{Dolgopol2} calculated the average scattering probabilities, 
$1/\tau^{\uparrow,\downarrow}$, corresponding to spin-up and 
spin-down electrons in a parallel magnetic field.  The longitudinal conductivity is 
the sum of contributions of spin-up and spin down-bands: $\sigma=\sigma^\uparrow + 
\sigma^\downarrow=n_s^\uparrow \mu^\uparrow + n_s^\downarrow \mu^\downarrow$, where 
$n_s^{\uparrow,\downarrow}=n_s/2(1 \pm \xi)$  is the density of the spin-up 
(spin-down) electrons and $\xi=H/H_{sat}, (H<H_{sat})$ is the degree of spin 
polarization of the 2D system.  Using their expression for the conductivity 
$\sigma(H)$ \cite{Dolgopol2} to calculate the mobilities 
$\mu^{\uparrow \downarrow}(H)$, the Hall coefficient $R_H$ was obtained 
as a function of parallel magnetic field $H$.

The calculated and measured Hall coefficients are shown as a function of parallel 
magnetic field in Fig. 2.  The experimental curve represents data obtained for three 
different densities: $1.91$, $2.47$, and $2.75 \times 10^{11}$ cm$^{-2}$.  A 
single-particle model which considers contributions 
from two bands of electrons that have different mobilities due to field-dependent 
screening predicts a Hall coefficient which varies substantially with magnetic 
field.  In contrast, the measured Hall coefficient is constant to within the 
$1-3$\% over 
the entire range of magnetic field up to well above $H_{sat}$.  This implies that 
the spin-up and spin-down electrons have mobilities that remain comparable at all 
fields, including very high magnetic field where the wave vectors of the spin-up
and spin-down electrons  $K_F^{\uparrow \downarrow}$ differ considerably.  Yet, we 
know that the average mobility of the electrons at $H=0$ depends strongly on the 
electron density and, therefore, on the value of $K_F$, especially 
near the transition.

One way to resolve this paradox is to consider the possibility that there are 
strong interactions between spin-up and spin-down electrons.  Frequent spin-flip 
scattering forces electrons to spend half the time in a spin-up state and 
half the time in a spin-down state.  Because of conservation of total 
\vbox{
\vspace{-0.15in}
\hbox{
\hspace{-0.28in} 
\epsfxsize 4.8in \epsfbox{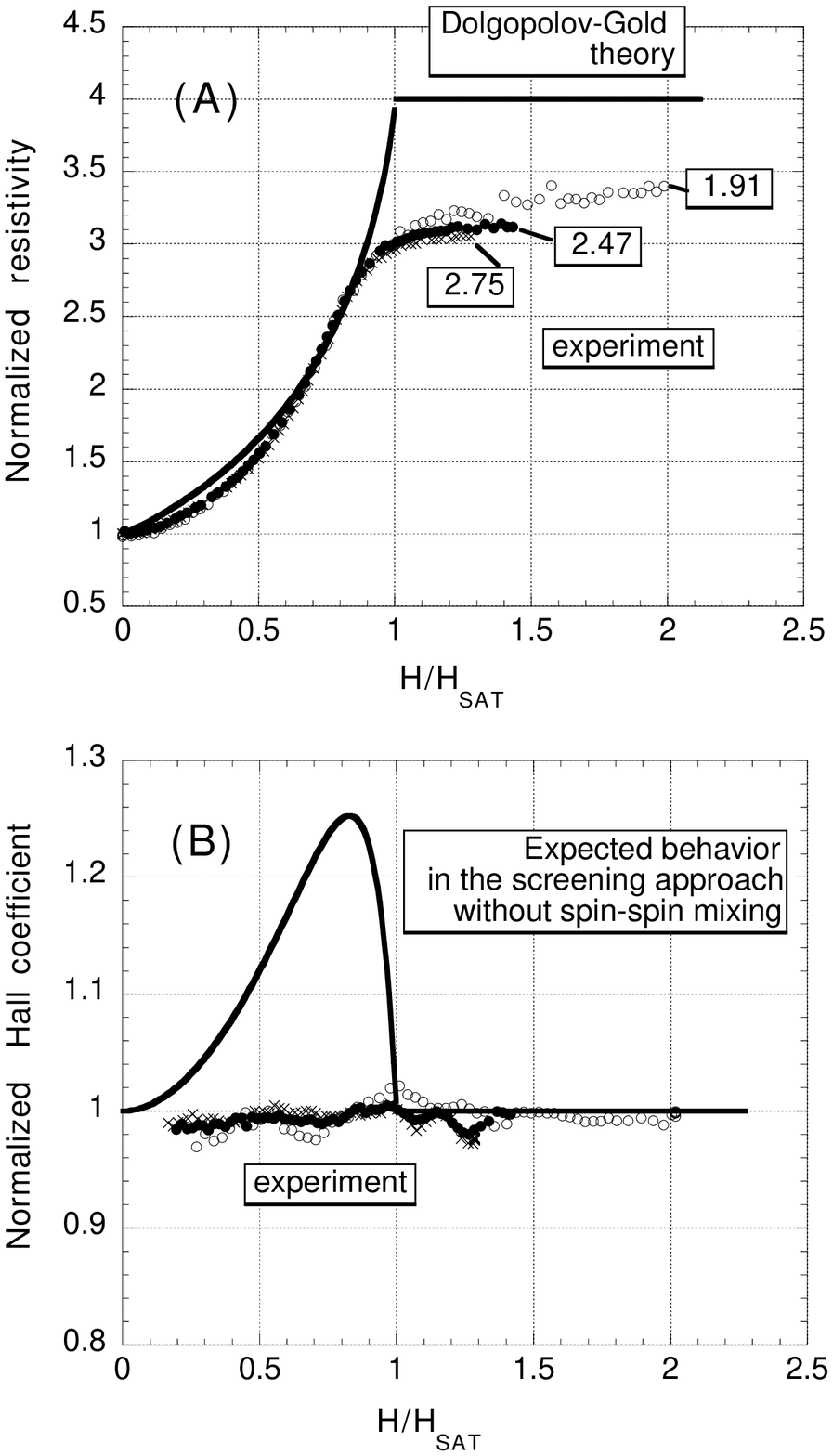} 
}
}
\refstepcounter{figure}
\parbox[b]{3.3in}{\baselineskip=12pt FIG.~\thefigure.
(A) The longitudinal resistance normalized to its zero-field value as a function of 
normalized in-plane magnetic field, $H/H_{sat}$.  The theoretical curve denotes the 
Dolgopolov-Gold theory for field-dependent screening of spin-up and spin-down 
carriers.  The three experimental curves are for densities $1.91$, $2.47$, and 
$2.75 \times 10^{11}$ cm$^{-2}$; the saturation field $H_{sat}$ is chosen to match 
the theoretical results of Dolgopolov and Gold.  (B) The Hall coefficient normalized 
to its zero-field value as a function of normalized in-plane magnetic field, 
$H/H_{sat}$.  The experimental curve is compared with the Hall coefficient expected 
from field-dependent screening (see text).
\vspace{0.10in}
}
\label{2}
momentum of the electron system, the strong scattering between spin-up and 
spin-down bands is not the same as the scattering responsible for the measured 
transport of the dilute 2D system.  The transport scattering probability is 
nevertheless strongly affected by the spin-flip electron-electron scattering, 
which results in electrons in either spin state having the same average transport 
scattering probability.  The strong mixing of spin-up and spin-down electron states 
may provide an explanation for our observation that the Hall coefficient is 
independent of in-plane magnetic field.  We note that we recently suggested that 
interactions between spin-up and spin-down electrons are also responsible for the 
unusual phase relation between the first and the second harmonics of the 
Shubnikov-de Haas oscillations in magnetic fields below saturation, $H<H_{sat}$ 
\cite{vitkalov}.

To summarize, measurements at low temperature of the longitudinal resistance and 
Hall coefficient as a function of in-plane magnetic field of silicon MOSFETs with 
high electron densities $n_s \geq 1.7 \times n_c$ (where $n_c$ is the critical 
density for the 
zero-field metal-insulator transition) are compared with expectations 
based on field-induced changes in the screening of spin-up and spin-down electrons.  
Fair agreement between theory and experiment is found for the longitudinal 
resistance at high densities well above $n_c$.  However, the Hall coefficient 
obtained by considering a two-band model of spin-up and spin-down electrons with 
different, field-dependent mobilities predicts a sizable dependence on in-plane 
magnetic field which is in clear contradiction with experiment.  We suggest that 
the Hall coefficient $R_H$ is independent of parallel magnetic field as a result of 
substantial mixing of spin-up and spin-down states due to strong 
electron-electron interactions in these dilute 
two-dimensional electron systems.

We are grateful to E. Abrahams, S. V. Kravchenko and V. T. Dolgopolov for valuable 
discussions and comments on our manuscript.  This work was supported by DOE grant 
No. DOE-FG02-84-ER45153.

\end{multicols}

\begin{references}



\bibitem{rmp} E. Abrahams, S. V. Kravchenko, and M. P. Sarachik, preprint
cond-mat/0006055 (2000) (to be published in Rev. Mod. Phys.); M. P. Sarachik and 
S. V. Kravchenko, Proc. Natl. Acad. Sci. {\bf 96}, 5900 (1999).
\bibitem{krav} S.\ V.\ Kravchenko, G.\ V.\ Kravchenko, J.\ E.\
Furneaux, V.\ M.\ Pudalov, and M.\ D'Iorio, Phys.\ Rev.\ B {\bf 50}, 8039
(1994); S.\ V.\ Kravchenko, W.\ E.\ Mason, G.\ E.\ Bowker,
J.\ E.\ Furneaux, V.\ M.\ Pudalov, and M.\ D'Iorio, Phys.\ Rev.\ B {\bf 51},
7038 (1995); S.\ V.\ Kravchenko,
D.\ Simonian, M.\ P.\ Sarachik, W.\ E.\ Mason, and J.\ E.\ Furneaux, Phys.\
Rev.\ Lett. {\bf 77}, 4938 (1996).
\bibitem{simonian} D.~Simonian, S.~V.~Kravchenko, M.~P.~Sarachik, and
V.~M.~Pudalov, Phys.\ Rev.\ Lett. {\bf 79}, 2304 (1997).
\bibitem{pudalov} V.~M.~Pudalov, G.~Brunthaler, A.~Prinz, and G.~Bauer,
Pisma Zh. Eksp. Teor. Fiz. {\bf 65}, 887 (1997)
[JETP Lett. {\bf 65}, 932 (1997)].
\bibitem{cambridge} M.~Y.~Simmons, A. R. Hamilton, M. Pepper, E. H.
Linfield, P. D. Rose,
D. A. Ritchie, A. K. Savchenko, and T. G. Griffiths, Phys.\ Rev.\ Lett. {\bf
80}, 1292 (1998).
\bibitem{yoon} J. Yoon, C. C. Li, D. Shahar, D. C. Tsui, and M. Shayegan,
Phys.\ Rev.\ Lett. {\bf 84}, 4421 (2000).
\bibitem{okamoto} T. Okamoto, K. Hosoya, S. Kawaji, and A. Yagi, Phys. Rev.Lett. 
{\bf 82}, 3875 (1999).
\bibitem{vitkalov} S. A. Vitkalov, H. Zheng, K. M. Mertes, M. P. Sarachik, and 
T. M. Klapwijk, Phys.\ Rev.\ Lett. {\bf 85}, 2164 (2000).
\bibitem{Dolgopol2} V. T. Dolgopolov and A. Gold, JETP Letters, {\bf 71},27, (2000).
\bibitem{ziman} J. M. Ziman, "Principles of the Theory of Solids", Cambridge 
University Press, (1972).
\bibitem{hairong} K. M. Mertes, H. Zheng, S. A. Vitkalov, M. P. Sarachik, and 
T. M. Klapwijk, preprint cond-mat/0006379 (2000).
\bibitem{Stern} F. Stern, Phys. Rev. Lett. {\bf 44}, 1469, (1980)
\bibitem{Ando1} T.Ando, J.Phys.Soc.Jpn. {\bf 51}, 3215, (1982).
\bibitem{DasSarma2} S. Das Sarma, Phys.Rev.Lett. {\bf 50}, 211, (1983).
\bibitem{Dolgop1} A. Gold and  V. T. Dolgopolov, Phys. Rev. {\bf B} {\bf 33}, 
1076 (1986). 
\end{references}
\end {document}